\documentclass{article}

\def\b{\begin{eqnarray}}
\def\e{\end{eqnarray}}
\def\n{\noindent}

\def\a{\alpha}
\def\p{\partial}

\def\f{\varphi}

\def\f{\phi}

\def\Box{\lower1pt \hbox{$\phantom{.}^{\mbox{\tiny \fbox{}}}$}}
\def\2{\frac{1}{A^2 - \phi}}

\def\phj{\phantom{j}}

\begin{document}

\begin{center}

{\huge\textbf{On the Field Equations of Kaluza's Theory \\}}
\vspace {10mm}
\noindent
{\large \bf Rossen I. Ivanov$^{\diamondsuit, \, \clubsuit, \, \ddagger}$ and Emil M. Prodanov$^{\spadesuit, \, \ast}$}
\vskip1cm \n
\hskip-.3cm
\begin{tabular}{c}
$\phantom{R^R}^\diamondsuit${\it School of Mathematics, Trinity College, University of Dublin, Ireland,} \\ \\
$\phantom{R^R}^\clubsuit$ {\it Institute for Nuclear Research and Nuclear Energy,} \\
{\it Bulgarian Academy of Sciences,}
\\ {\it 72 Tsarigradsko chaussee, Sofia -- 1784, Bulgaria} \\ \\
$\phantom{R^R}^\spadesuit${\it School of Physical Sciences, Dublin City University, Glasnevin, Ireland} \\ \\ \\
$\phantom{R^R}^\ddagger${\it ivanovr@tcd.ie} \\ \\
$\phantom{R^R}^\ast${\it prodanov@physics.dcu.ie}
\end{tabular}
\vskip1cm
\end{center}

\vskip2cm
\begin{abstract}
The field equations of the original Kaluza's theory are analyzed and it is shown that they lead to modification
of Einstein's equations. The appearing extra energy-momentum tensor is studied and an example is given where
this extra energy-momentum tensor is shown to allow four-dimensional Schwarzschild geometry to accommodate
electrostatics. Such deviation from Reissner--Nordstr\o m geometry can account for the interpretation of
Schwarzschild geometry as resulting not from mass only, but from the combined effects of mass and electric
charge, even electric charge alone.

\end{abstract}

\vfill

$\phantom{EMP}$ \\ $\phantom{EMP}$ \\ $\phantom{EMP}$ \\ $\phantom{EMP}$

\newpage

\section{Introduction}
Kaluza--Klein's theory has been an area of extensive research for almost a century \cite{chodos}. Kaluza's
original theory \cite{kaluza} seems unattractive because of the apparent lack of gauge invariance (it is Klein's
later modification \cite{klein} which proposes the gauge invariant version of the theory). The two theories,
however, are intrinsically related and dual --- these are the slicing and threading decomposition of a
five-dimensional spacetime \cite{boersma}. Therefore, one would expect that the physics in either of these
two pictures should somehow induce the physics in the other one. The physics behind the field equations of
the original Kaluza's theory is compared to that of Klein's theory, as analyzed by Jordan and Thiry \cite{th}.
It is shown that in Kaluza's model, gauge invariance of the ``electromagnetic'' fields is actually transferred as a
gauge freedom to fix the dilaton field (which models Newton's constant) as needed by experiments. This is the
crucial difference between these two dual theories: Klein's theory fails to achieve a constant solution for the
dilaton since an unwanted constraint for the Maxwell's tensor appears; Kaluza's theory sacrifices the gauge
freedom of the electromagnetic fields --- the field equations of the original Kaluza's theory contain gauge-related
(or {\it ghost}) terms. These terms are studied in the case of a constant dilaton. We suggest that the extra
energy-momentum tensor could contribute to the tackling of the dark matter problem. We also give an example which
shows that the ghost matter terms allow Reissner--Nordstr\o m geometry to deviate into Schwarzschild geometry.
This leads to interpretation of Schwarzschild geometry as arising not from mass only, but from the combined
effects of mass and electric charge or even from electric charge alone.

\section{Kaluza--Klein Duality}

In this section we give an introduction to the {\it slicing--threading duality} \cite{boersma}.
The original Kaluza's theory \cite{kaluza} and Klein's subsequent modification \cite{klein} are dual theories.
Namely, this is a duality between decomposition of the five-dimensional spacetime with dimension four surfaces
(slicing) and the decomposition of the five-dimensional spacetime with co-dimension four surfaces (threading). \\
We will explain the decomposition of a bundle metric in terms of the metrics
of the base space and the fibre space --- see \cite{boersma} for the details. Let $\mathcal{M}$ be an
$(m+n)$--dimensional fibre bundle with $m$--dimensional base space $\mathcal{B}$ (with local coordinates
$x^1, \,  x^2, \, \ldots \, , x^m$), $n$--dimensional fibre $\mathcal{F}$ (with local coordinates
$y^1, \,  y^2, \, \ldots \, , y^n$) and projection $\pi_1: \, \mathcal{M} \rightarrow \mathcal{B}$. The
projection $\pi_1$ defines a local trivialization for each neighbourhood $U$ of the point $x \in \mathcal{B} \, $,
$\, \, \pi_1^{-1}(U) \simeq U \times \mathcal{F}$ with local coordinates $(x^i,y^a) \, $ (in this section only,
the indexes $i, j$ run from 1 to $m$, the indexes $a, b$ --- from 1 to $n$ and Greek indexes --- from 1 to $m + n$).
Let $G$ be a metric, defined on $T \mathcal{M} \otimes T \mathcal{M}$. At each point $P$ of $\mathcal{M}$ there
exists a natural subspace, called the vertical subspace $\mathcal{V}_P \subset T_P \, \mathcal{M}$ with basis
$\{ \partial / \partial y^a\}$. Let us define a horizontal space $\mathcal{H}_P$, complementing
$\mathcal{V}_P$: $T_P \mathcal{M} = \mathcal{V}_P \oplus \mathcal{H}_P \, $, such that the two are orthogonal
with respect to $G$, i.e. if $H_1 \, , \, H_2 \, , \, \ldots \, , H_n$ is the basis of $ \mathcal{H}_P$, then:
\b
\label{eq1}
G \Bigl( H_i \, , \,\,  \frac{\partial}{\partial y^{a}} \Bigr) = 0 \, ,
\e
where $H_i = \partial / \partial x^i - \Gamma^a_i \, \, \partial / \partial y^a \, $ is a
{\it horizontal lift} of $\partial / \partial x^i$. We will use the following shorthand notations:
$G_{ij} \equiv G(\partial / \partial x^i \, , \, \partial / \partial x^j)$,
$G_{ia} \equiv G(\partial / \partial x^i \, , \, \partial / \partial y^a)$,
$G_{ab} \equiv G(\partial / \partial y^a \, , \, \partial / \partial y^b)$. \\
The metric $g$ on $T \mathcal{B} \otimes T \mathcal{B}$ can be defined as:
\b
\label{eq2}
g \Bigl(\frac{\partial}{\partial x^{i}} \, , \, \frac{\partial}{\partial x^{j}} \Bigr)
\, \equiv \, \, G(H_{i} \, , \, H_{j})
\e
and, accordingly, we will use the shorthand notation $g_{ij} \equiv g (\partial / \partial x^i \, , \,
\partial / \partial x^j)$. \\
The metric $h$ on $T \mathcal{F} \otimes T \mathcal{F}$ will be defined simply as a pull-back metric of $G$:
\b
\label{eq3}
h \Bigl( \frac{\partial}{\partial y^{a}} \, , \,
\frac{\partial}{\partial y^{b}} \Bigr) \, \equiv \, \,
G \Bigl( \frac{\partial}{\partial y^{a}} \, , \, \frac{\partial}{\partial y^{b}} \Bigr) \, ,
\e
or $h_{ab}\equiv G_{ab}$ for short. \\
Now from (\ref{eq1}) and (\ref{eq3}), it follows that $G_{ia} = \Gamma^b_i \, h_{ab}$ and using (\ref{eq2}) and
(\ref{eq3}), we obtain the decomposition in the described bases:
\b
\label{eq4}
G_{\mu \nu} \, = \,
\left( \begin{array}{c|c}
\\
G_{ij} &  G_{bj} \\ \\
\hline
\\
G_{ia} & G_{ab} \\ \\
\end{array}  \right) \, = \,
\left( \begin{array}{c|c}
\\
g_{ij} + \Gamma^a_i \Gamma^b_j \, h_{ab} & \Gamma^b_j \, h_{ab}  \\ \\
\hline
\\
\Gamma^{a^{\phantom{A^A}}}_{i} \!\!\!\!\!\!\!h_{ab} & \phantom{Qwer} h_{ab} \phantom{Qwer}\\ \\
\end{array}  \right) \, .
\e
\vskip.5cm
\noindent
\underline{Example 1}: $m=1$  (slicing). \\
In this case, $\mathcal{F}$ is still an $n$--dimensional hypersurface, $\mathcal{B} = \mathrm{R}^1 \, ,
\, \Gamma^a_1 \equiv B^a$ is an $n$-vector, and $g = g_{11}$ is a scalar. Defining the new scalar
$\phi \equiv g_{11} + h_{ab} \, B^a \, B^b \, ,$ we obtain the $(1+n)$--decomposition, known as {\it slicing}:
\b
\label{eq5}
G_{\mu \nu} =
\left(
\begin{array}{c|ccc}
\phi & & & B_a \phantom{Qwert} \\
\hline
 & &  & \\
B_a & & & h_{ab} \phantom{Qwert} \\ \\
\end{array}
\right) \, .
\e
For $n = 4$,  one can immediately identify here the original Kaluza's metric \cite{kaluza}.
\vskip.5cm
\noindent
\underline{Example 2}: $n=1$ (threading). \\
In this case, $\mathcal{B}$ is an $m$--dimensional hypersurface, $\mathcal{F} = \mathrm{R}^1 \, , \,
\Gamma^{1}_{i} \equiv A_{i}$ is an $m$-vector, and $h = h_{11}$ is a scalar which we take as $V^2$.
The $(m+1)$--decomposition, known as {\it threading}, is:
\b
\label{eq6}
G_{\mu \nu} =
\left(
\begin{array}{ccc|c}
& & & \\
& g_{ij} +V^{2} A_{i}A_{j}& & V^{2}A_{i}  \\
& &  & \\
\cline{1-4}
& V^{2}A^{^{\phantom{A^A}}}_{i} \!\!\!\!\!\!\! & & V^{2} \\
\end{array}
\right) \, .
\e
For $m = 4$, this reduces to the Klein's metric \cite{klein}. \\
One should also note that the Kaluza's metric (\ref{eq5}) has the same form as the inverse of Klein's metric (\ref{eq6}).

\section{Spacetime Structure}

We would like now to review the spacetime structure of Kaluza--Klein theory and prepare the set-up for our
further analysis. \\
Kaluza's original theory is a unitary representation of gravitation and electromagnetism. The stage for this theory
is a five-dimensional spacetime foliated by a one-parameter family of four-dimensional hypersurfaces, each of which
could be interpreted as a ``particular'' universe. The geometry of the five-dimensional spacetime arises from the
geometries of the four-dimensional ``slices''. As suggested by Kaluza in his original paper, it is quite natural
to demand that the physical quantities of the four-dimensional world be independent on the extra dimension
(denoted by $x$ in Kaluza's theory).
The five-dimensional Kaluza's metric is simply the (1+4)--decomposition (\ref{eq5}) and the corresponding interval is:
\b
\label{tua}
d s^2 = h_{ij} (dy^i + B^i dx)(dy^j + B^j dx) + \frac{1}{N^2} dx^2 \, ,
\e
where $N^{-2} = \f -B^2$. From now on, Latin indexes run from 1 to 4 and lowering and raising of Latin indexes is
made with $h_{ij}$ and its inverse $h^{ij}$, respectively. \\
The slicing lapse function is $N^{-1}$, while the slicing shift vector field is given by $B^i$. \\
The tensor $h_{ij}$ was identified as the metric of our four-dimensional world (it appears naturally as a
slicing metric), and the vector $B_i$ --- as the electromagnetic potential. The antisymmetrical covariant derivatives
of $B_i$ with respect to $y^j$, namely $F_{ij} = \nabla_i B_j - \nabla_j B_i = \p_i B_j - \p_j B_i \, ,$ were identified
as components of the Maxwell tensor. The scalar $\f$ was left uninterpreted. \\
The symmetrical covariant derivatives of $B_i \, ,$ that is $\Sigma_{ij} = \nabla_i B_j + \nabla_j B_i \, , $ also enter
the field equations. For the consistent interpretation of Kaluza's theory as a five-dimensional theory of gravitation
and electromagnetism, one has to make sure that the symmetrical derivatives $\Sigma_{ij} = \nabla_i B_j + \nabla_j B_i$
vanish in the whole space. This is the so called "cylinder condition". Mathematically, this condition is achieved by
requesting that the five-dimensional interval (\ref{tua}) is invariant under the transformation \cite{eb}:
\b
y^\mu \,\, \rightarrow \,\,  y^\mu + \epsilon \, B^\mu \, ,
\e
where $y^0 = x \, , \, B^0 = \f $ and $\epsilon$ is an infinitesimal parameter (Greek indexes from now on
run from 0 to 4). \\
Kaluza's theory imposes another property on the vector field $B^i$ --- the lines to which $B^i$ are tangents (we will call
these threads) must be geodesics \cite{eb}. This results in the fact that the norm of the vector $B$ is constant in
the whole space: $B^i \nabla_j B_i = 0$. \\
The cylinder condition is rather artificial --- as argued by Einstein and Bergmann \cite{eb}, it is not really a
step forward to introduce a five-dimensional metric and a vector for which arbitrary restrictions are assumed,
instead of having a four-dimensional metric and a four-vector --- the real step forward, according to Einstein and
Bergmann, was the introduction of the five-dimensional space alone. Additionally, in Kaluza's original theory, the vector
$B_i$ is not defined up to an additive gradient of an arbitrary function, namely, Kaluza's theory lacks gauge
invariance. Einstein and Bergmann generalized Kaluza's theory as follows \cite{eb}. They considered a four-dimensional
surface, cutting each of the geodesics to which $B^i$ are tangents (the threads) once and only once.
The four coordinates, introduced on this surface are $y^i$  and the extra coordinate $x$ is assumed constant on
this surface. As shown by Einstein and Bergmann, this coordinate system is preserved by only two
coordinate transformations: \\
\underline{(i) Four-transformation:}
\b
\label{four}
y^i & \rightarrow & y'^i \, (y^j) \, , \nonumber \\
x & \rightarrow & x \, ,
\e
\underline{(ii) Cut-transformation:}
\b
\label{cut}
y^i & \rightarrow & y^i \, , \nonumber \\
x & \rightarrow & x + f(y^i) \, .
\e
The four-transformation (\ref{four}) is a transformation from one thread to another while staying on the
same four-dimensional slice. This transformation is the only natural transformation for a five-dimensional spacetime
foliated by four-dimensional surfaces (slicing decomposition, or the original Kaluza's theory). The cut-transformation
(\ref{cut}) is a transformation from one four-dimensional slice to another while staying on the same thread.
This is the only natural transformation for a five-dimensional spacetime foliated by surfaces of co-dimension four,
i.e. a congruence of threads (threading decomposition, or Klein's theory). \\
Under the four-transformation (\ref{four}), the four-dimensional metric $h_{ij}$ transforms as a tensor and $B_i$
transforms as a vector. However, this is not the case under the cut-transformation (gauge transformation) (\ref{cut}):
\b
\label{gage}
B_i & \rightarrow & B_i - \partial_i f \, , \\
h_{ij} & \rightarrow & h_{ij} - B_i \partial_j f - B_j \partial_i f + (\partial_i f)(\partial_j f) \, .
\e
However,
\b
\label{ejgo}
g_{ij} = h_{ij} - B_i B_i
\e
are invariant under the cut-transformation \cite{eb}. Apparently, even though that $B_i$ does not transform
as a four-vector, the quantity $F_{ij} = \nabla_i B_j - \nabla_j B_i = \p_i B_j - \p_j B_i$ is invariant under
the cut-transformation. In view of the transformation law (\ref{gage}), the vector $B_i$ is defined up to an additive
gradient of an arbitrary function --- exactly as an electromagnetic potential. The invariant $F_{ij}$ is then
Maxwell's tensor. In view of its gauge invariance, $g_{ij}$ can be identified as a metric tensor.
However, a surface whose metric is $g_{ij}$ does not necessarily exist. The tensor $g_{ij}$ gives the proper
distance $\sqrt{g_{ij} d y^i d y^j}$ between two threads passing through points $P_1(y^i, x)$ and
$P_2(y^i + d y^i, x + d x)$ (the distance between points $P_1$ and $P_2$ is given by the interval $d \sigma$) \cite{boersma}. \\
In addition, Einstein and Bergmann, in their quest for simplification of the basic geometric assumptions, proposed to
drop the artificial cylinder condition and, instead, introduce a periodicity condition: space is periodic in the extra
direction \cite{eb}. These are the ingredients of the theory generally referred today as Kaluza--Klein theory and,
without doubt, its precise mathematical formulation must be attributed to Einstein and Bergmann. We will refer to this
picture as Klein's theory. \\
There is an extremely important observation due here. If $\psi$ is some scalar, then $\partial_i \psi$ is not invariant
under the cut-transformation --- the only transformation allowed in Klein's theory. It is the starry derivative,
\b
\partial^\ast_i = \partial^{\phantom{\ast}}_i - B^{\phantom{\ast}}_i \partial^{\phantom{\ast}}_x \, ,
\e
that is invariant under the cut-transformation \cite{eb}. To build tensor analysis with respect to the
 cut-transformation, namely to build tensor analysis in Klein's theory, one has to invoke the starry derivative.
Therefore, the Riemann curvature tensor
\b
R^i_{\phantom{i}jkl} = \partial_l \Gamma^i_{jk} - \partial_k \Gamma^i_{jl}
- \Gamma^i_{km} \Gamma^m_{jl} + \Gamma^i_{lm} \Gamma^m_{jk} \, ,
\e
must be replaced by the tensor \cite{eb}:
\b
\label{zed}
Z^i_{\phantom{i}jkl} = R^i_{\phantom{i}jkl} - B_l \, \partial_x \Gamma^i_{jk} +
B_k \, \partial_x \Gamma^i_{jl} \, .
\e
Today, this tensor is referred to as Zelmanov curvature tensor \cite{zel}. \\
Replacing the Riemann curvature tensor by the tensor (\ref{zed}) corresponds to a surface forming mechanism --- in
Klein's theory the four-dimensional world is not naturally formed as it is in Kaluza's original theory. In Klein's
theory, the congruence of $x$-like curves could be interpreted as the ``world lines'' of a family of observers
and the ``rest frames'' of these observers do not necessarily form a surface. Requesting independence on the
extra dimension means that $Z^i_{\phantom{i}jkl} = R^i_{\phantom{i}jkl}$ and that the hypersurfaces for different
$x$ are "smeared" into one (which is not a gross injustice under the condition of periodicity along the extra
dimension). \\
We are now in a position to define the set-up for our analysis. The field equations of Klein's theory are well known
(for the sake of completeness, we dedicate the next section to them). It is interesting to study the field equations
of the original Kaluza's theory in view of the duality between the two models. Of course, the additional conditions
imposed, namely cylinder condition versus periodicity condition are not related under the Kaluza--Klein
(slicing--threading) duality and we will study the full field equations of Kaluza's theory without imposing the
cylinder condition. Mathematically, imposing the cylinder condition, together with imposing independence on the extra
dimension $x$, amounts to requesting zero extrinsic curvature. We will keep the extrinsic curvature and study the field
equations as found in \cite{ip} and \cite{ws}. We will disregard dependence on the extra dimension. We will show that
the resulting field equations are plausible generalizations of Einstein's and Maxwell's equations, we will study the
transfer of gauge invariance from the fields $B_i$ to the dilaton (which models Newton's constant) and we will give an
illustration with a particular example for constant dilaton.

\section{Field Equations of Klein's Theory}

Klein's model has the (4+1) metric (\ref{eq6}). The five-dimensional interval is:
\b
d s^2 = g_{ij} dx^i dx^j + V^2 (A_i dx^i + d y )^2 \, .
\e
Here $x^i$ are the four-dimensional coordinates and $y$ labels the extra dimension. The field $V^2$ is the threading
lapse function, while $A_i dx^i$ is the threading shift one-form. The fields $g_{ij}$ are the components of the
threading metric. \\
The field equations $\mathcal{R}_{\mu \nu}^{(5)} = 0$ are \cite{th}:
\b
\label{t1}
\mathcal{R}_{ij} - \frac{1}{2} g_{ij} \mathcal{R}
& = & \frac{V^2}{2} \mathcal{T}_{ij}^{\mbox{\tiny EM}} - \frac{1}{V} \Bigl[ D_i (\p_j V) - g_{ij} \Box V \Bigr] \, , \\
\label{t2}
D_i \tilde{F}^{ij} & = & -3 \frac{\p_i V}{V}  \tilde{F}^{ij} \, , \\
\label{t3}
\Box V & = & \frac{V^3}{4} \tilde{F}_{ij} \tilde{F}^{ij} \, ,
\e
where $\mathcal{R}_{\mu \nu}^{(5)}$ is the five-dimensional Ricci tensor, $\mathcal{R}_{ij}$ is the four-dimensional
Ricci tensor, $\mathcal{R}$ is the four-dimensional scalar curvature, $D_i$ is the four-dimensional covariant
derivative, $\Box = g^{ik} D_i D_k$,  $\tilde{F}_{ij} = \p_i A_j - \p_j A_i$ is the Maxwell tensor, and the derivatives are
with respect to $x^i$. In mostly plus metric, the electromagnetic energy-momentum tensor is:
\b
\mathcal{T}_{ij}^{\mbox{\tiny EM}}  =  \tilde{F}_{il} \tilde{F}^{\phj l}_j - \frac{1}{4} g_{ij}
\tilde{F}_{kl} \tilde{F}^{kl} \, .
\e
One notes from equation (\ref{t1}) that Newton's constant $G_N$ is expressed as a dynamical field:
\b
\frac{V^2}{2} = \kappa = \frac{8 \pi G_N}{c^4} \, .
\e
A constant solution for $V$ (and, consequently, for the Newton's constant) reduces equations (\ref{t1})
and (\ref{t2}) to the usual Einstein and Maxwell equations. \\
However, in this case equation (\ref{t3}) reduces to $\tilde{F}^2 = \tilde{F}_{ij} \tilde{F}^{ij}
= 2(c^2 B^2 - E^2) = 0$, that is the square of the electric field must be equal to $c$ times the square of the
magnetic field. While this is indeed the case for plane electromagnetic waves for instance, there is no physical
mechanism which will require $\tilde{F}^2 = 0$. In this sense, $\tilde{F}^2 = 0$ is an unphysical condition.

\section{Field Equations of Kaluza's Theory}

For Kaluza's model (\ref{tua}), the lack of appropriate gauge invariance for the fields $B^i$, which we, nevertheless,
will identify with the electromagnetic potentials, is transformed as a gauge degree of freedom for the dilaton $N$.
As a result we end up with a gauge-fixed electrodynamics, but we are free to fix the value of $N$ as needed by the
model, including $N = \mbox{const}$.  This corresponds to the electromagnetic gauge freedom in the model of
Klein \cite{klein} (see also the equations of Jordan--Thiry \cite{th}). \\
Let us denote the five-dimensional Ricci tensor by $R_{\mu \nu}^{(5)}$ , the four-dimensional Ricci tensor by
$r_{ij}$ and the four-dimensional scalar curvature by $r$. As before, we use $\nabla_i$ for the four-dimensional
covariant derivative with respect to $y^i$ and $F_{ij} = \p_i B_j - \p_j B_i$ for the Maxwell electromagnetic tensor.
The extrinsic curvature is $\pi_{ij} = - (N/2)(\nabla_i B_j + \nabla_j B_i)$. Then the field equations
$R_{\mu \nu}^{(5)} = 0$, determined in \cite{ip}, are (see also \cite{ws} for their form in terms of the extrinsic
curvature $\pi_{ij}$):
\b
\label{einstein}
& & r_{ij} - \frac{1}{2} h_{ij}r  =  \frac{N^2}{2} T_{ij} \, , \\
\label{maxgen}
& & \nabla_i F^{ij} = - 2 B_i r^{ij} + \frac{2}{N^2} (\pi^{ij} - \pi^k_k h^{ij}) \p_i N \, , \\
\label{div}
& & \nabla_i (B_j \pi^{ij} + \frac{1}{N^2} \nabla^i N) = 0 \, .
\e
Here, $x$-dependent terms are dropped out. \\
The dilaton $N$ is related to the Newton's constant $G_N$ via \cite{ip}:
\b
\frac{N^2}{2} = \kappa = \frac{8 \pi G_N}{c^4} \, .
\e
The energy-momentum tensor $T_{ij}$ appearing in equation (\ref{einstein}) is given by \cite{ip}:
\b
\label{te}
T_{ij} = T^{\mbox{\tiny EM}}_{ij} \, + \, \nabla^k \Psi_{ijk} \, + \, \nabla^k \Theta_{ijk} \,
+ \, C_{ij} \, + \, D_{ij} \, ,
\e
where:
\b
T^{\mbox{\tiny EM}}_{ij} & = & F_{ik} F^{\phj k}_j - \frac{1}{4} h_{ij} F_{kl} F^{kl} \, , \\ \nonumber \\
\label{24}
\Psi_{ijk} & = & B_k \nabla_j B_i - B_j \nabla_k B_i + B_i F_{jk} \, , \\ \nonumber \\
\Theta_{ijk} & = & \nabla_i (B_k B_j) + h_{ij}(B^l \nabla_k B_l - B_k \nabla_l B^l)  \, ,  \\ \nonumber \\
C_{ij} & = & h_{ij} B^k B^l r_{kl} - 2 B^k B_j r_{ik} - 2 B^k B_i r_{jk} \, , \\ \nonumber \\
D_{ij} & = & \frac{2}{N} \nabla_i \nabla_j \frac{1}{N} - \frac{2}{N^2} \pi^k_k (B_i \p_j N + B_j \p_i N)
\nonumber \\
& & + \frac{2}{N^2} \Bigl[ - B^k \pi_{ij} + B_i \pi^k_j + B_j \pi^k_i - h_{ij}
    (B^l \pi^k_l - B^k \pi^l_l)\Bigr] \p_k N \, .
\e
Let us now consider the case when the dilaton $N$ is constant, i.e. Newton's constant is indeed a constant. Kaluza's
equations then reduce to:
\b
\label{e1}
(r_{ij} - \frac{1}{2} h_{ij}r) - \frac{N^2}{2} T^{\mbox{\tiny G}}_{ij} & = & \frac{N^2}{2}
T^{\mbox{\tiny EM}}_{ij} + \frac{N^2}{2} C_{ij} \, , \\
\label{e2}
\nabla_i F^{ij} & = & -2 B_i r^{ij}  \, , \\
\label{e3}
\nabla_i (B_j \pi^{ij}) & = & 0 \, ,
\e
Here one can immediately recognize (\ref{e2}) as a generalization of Maxwell's equations. The right-hand-side
describes interaction between the electromagnetic fields and gravitation. The first of these equations,
(\ref{e1}), is a modified Einstein's equation. Apart from the Maxwell's energy-momentum tensor
$T^{\mbox{\tiny EM}}_{ij}$, on the right-hand-side we have the tensor $C_{ij}$ which describes interaction between
the electromagnetic fields and gravitation. On the left hand-side we have the modified Einstein's tensor
$r_{ij} - \frac{1}{2} h_{ij}r - (N^2 / 2) T^{\mbox{\tiny G}}_{ij}$. We will give interpretation of the
extra term $T^{\mbox{\tiny G}}_{ij}$ as a ghost energy-momentum tensor and we will comment on the possible implications
of such tensor. The last equation, (\ref{e3}), is an equation for $B_i$, additional to Maxwell's equations (\ref{e2})
and can be interpreted as a gauge-fixing equation for the electromagnetic potentials (in Kaluza's theory the
electromagnetic potentials $B_i$ are gauge-fixed). \\
The parameter $\kappa = N^2 / 2 = 8 \pi G_N / c^4$ is small. Then, from (\ref{einstein}), we see that $r_{ij}$ is
of order of $\kappa$. Using Landau symbols, $r_{ij} = \mathcal{O}(\kappa)$. Thus, $B_i r^{ij} = \mathcal{O} (\kappa)$
and $\frac{N^2}{2} C_{ij} = \mathcal{O} (\kappa^2)$. Note that the scale of $B_i$ cannot be increased with a gauge
transformation and thus change the order of the interaction terms $B_i r^{ij}$ or $\frac{N^2}{2} C_{ij}$. \\
Taking covariant derivative from equation (\ref{e1}), and using the contracted Bianchi identities,
$\nabla^j (r_{ij} - \frac{1}{2} h_{ij}r) = 0$,  yields:
\b
\label{im}
\nabla^j T^{\mbox{\tiny EM}}_{ij} +  \nabla^j  \nabla^k \Theta_{ijk} + \nabla^j \nabla^k \Psi_{ijk}
+ \nabla^j  C_{ij}  = 0 \, .
\e
The meaning of the tensors involved is best understood if we consider the leading order of $\kappa$. For the
Maxwell energy-momentum tensor we have:
\b
\label{eqm1}
\nabla^j T^{\mbox{\tiny EM}}_{ij} = F_{ik} \nabla_j F^{jk} = - 2 B_j F_{ik} r^{jk} = \mathcal{O} (\kappa) \, ,
\e
due to (\ref{e2}). Equation (\ref{eqm1}) is the conservation law for the energy and momentum resulting from
Maxwell's equations (\ref{e2}). \\
The tensor $\nabla^k \Theta_{ijk}$ satisfies:
\b
\label{eqm2}
\nabla^j \nabla^k \Theta_{ijk} \! = \!
- \frac{2}{N} \nabla_j \nabla_i (B_k \pi^{ik}) + \mathcal{O} (\kappa) = 0 + \mathcal{O} (\kappa) ,
\e
in view of (\ref{e3}). We now note that Maxwell's equations (\ref{e2}) are to $T^{\mbox{\tiny EM}}_{ij}$ as the
gauge-fixing equation (\ref{e3}) is to $\nabla^k \Theta_{ijk}$ (i.e. each of equations (\ref{e2}) and (\ref{e3}),
in turn, guarantees the vanishing of the covariant derivatives of the tensors $T^{\mbox{\tiny EM}}_{ij}$ and
$\nabla^k \Theta_{ijk}$, respectively). And since (\ref{e3}) is the gauge-fixing equation, we will interpret
$\nabla^k \Theta_{ijk}$ as a ghost energy-momentum tensor. \\
Considering the remaining term, $\nabla^k \Psi_{ijk}$, we see that it satisfies:
\b
\label{eqm3}
\nabla^j \nabla^k \Psi_{ijk} & = & \frac{1}{2} (\nabla^j \nabla^k + \nabla^k \nabla^j) \Psi_{ijk}
+ \frac{1}{2} [\nabla^j, \nabla^k] \Psi_{ijk} = 0 + \mathcal{O} (\kappa)
\e
in view of the antisymmetry $\Psi_{ijk} = - \Psi_{ikj}$. (The second term is of order $\mathcal{O}(\kappa)$.) Note
that, due to (\ref{im}), $\nabla^j C_{ij}$ exactly compensates the sum of the terms of order $\mathcal{O}(\kappa)$,
which we put aside in equations (\ref{eqm1})--(\ref{eqm3}). The tensor $\nabla^k \Psi_{ijk}$ does not describe any
dynamics (in leading order). Its only purpose is to make the linear combination $\nabla^k \Theta_{ijk} +
\nabla^k \Psi_{ijk}$ symmetric under exchange of indexes $i$ and $j$. Therefore, the full ghost energy-momentum
tensor is given by the Belinfante tensor:
\b
\label{gho}
T^{\mbox{\tiny G}}_{ij} =  \nabla^k (\Theta_{ijk} + \Psi_{ijk}) \, .
\e
Form equations (\ref{eqm2}) and (\ref{eqm3}) it follows that
\b
\nabla^j T^{\mbox{\tiny G}}_{ij} =  0 \, ,
\e
which represents the ghost conservation law for equation (\ref{e3}), analogical to the matter conservation
law (\ref{eqm1}) for equation (\ref{e2}).

\section{Applications}

The observed flat galaxy rotation curves is important evidence for the existence of a large fraction of dark matter
in the Universe in the framework of Newton's and Einstein's theories of gravitation. Over the last decades however,
alternative scenarios which modify Newton's and Einstein's theories have gained momentum. The first attempt in this
direction was made by Einstein himself in search for a static solution to Einstein's equations:
\b
\label{l1}
E_{ij} + \Lambda g_{ij} = \kappa T_{ij} \, ,
\e
where $E_{ij} = r_{ij} \, - \, \frac{1}{2} h_{ij} \, r \, $ is the Einstein's tensor, $T_{ij}$ is the material energy
tensor and $\Lambda$ is the cosmological constant. \\
Later on, in 1948, Bondi, Gold and Hoyle \cite{hoyle} proposed the steady-state cosmological model by the manual
introduction of an extra tensor in Einstein's equations which continuously generates matter that compensates the
Universe's expansion and makes the Universe look perpetual:
\b
\label{l2}
E_{ij} + C_{ij} = \kappa T_{ij} \, .
\e
The newly introduced tensor $C_{ij}$ is defined as the covariant derivative of some constant vector field:
$C_{ij} = \nabla_i C_j$. This tensor plays a role similar to that of the cosmological constant and, as the cosmological
constant, substantially changes the physical picture. \\
Until the discovery of the cosmic microwave background radiation, the steady-state cosmology was a viable alternative
to the Big Bang model. Recently, it was revived as quasi-steady-state theory \cite{quasi}. In the framework of the
(quasi)-steady-state theory, the dark matter problem takes on a different complexion --- there is no restriction like
$\Omega = 1$ in this cosmology and so the dark matter component need not be very high \cite{narlikar}. The extent of
dark matter has to be estimated from improved observations. Even though that our model has nothing to do with the
(quasi)-steady-state theory, exactly the same issue surfaces here --- we have shown (rather than introduced) a very
similar modification of Einstein's equations (\ref{e1})which, again, cannot lead to the restriction $\Omega = 1$. \\
Other reasonably simple modifications of gravity which describe galaxy rotation curves quite well, {\it without
requiring the existence of dark matter at all}, have been proposed --- see the topical review \cite{burgess} and the
references therein. \\
Therefore, a possible scenario for the avoidance of the dark matter problem (by modification of gravity) is to perform
a {\it cosmological gauge transformation} of the Einstein's tensor $E_{ij} = r_{ij} \, - \, \frac{1}{2} h_{ij} \, r$ by
an extra (ghost) energy-momentum tensor:
\b
\label{moral}
E_{ij} \to E_{ij} - \frac{N^2}{2} \nabla^k (\Theta_{ijk} + \Psi_{ijk}) \, .
\e
Such transformation is a continuation along the line of modifications of the type (\ref{l1}) and (\ref{l2}) of
Einstein's theory and, as it substantially changes the physical picture by the introduction of mass and energy, it
could account for the description of the missing matter and energy in the Universe. \\
Interestingly, it was shown by Arkani-Hamed {\it et al.} \cite{ah} that ghost condensate may contribute to both the
dark matter and the dark energy.

\vskip.2cm
\n
Returning to the field equations of Kaluza's theory, we consider, as an example, the following five-dimensional metric:
\b
\label{we}
d s^2 & = & - \, (1 - \frac{\a}{r}) dt^2 - 2 \beta(1 - \frac{\a}{r})dt dx + (1 - \frac{\a}{r})^{-1} dr^2
\nonumber \\ & & \hskip.3cm + \,\,  r^2 d \theta^2
+ r^2 \sin^2 \theta \, d\varphi^2 + \Bigl[1 - (1 - \frac{\a}{r})\beta^2 \Bigr] dx^2
\e
which is a solution to the five-dimensional vacuum Einstein's equation $R^{(5)}_{\mu \nu} = 0$. This metric corresponds
to $r_{ij} = 0$ and $N = 1$. One can identify
\b
B_i = - \beta(1 - \frac{\a}{r}) \delta_{i0}
\e
as an "electrostatic" potential generated by charge
\b
q = \alpha \beta \, .
\e
The metric (\ref{we}) is also a solution to the following field equations:
\b
\label{ne1}
T^{\mbox{\tiny EM}}_{ij} + T^{\mbox{\tiny G}}_{ij} & = & 0 , \\
\label{ne2}
\nabla_i F^{ij} & = & 0 \, , \\
\label{ne3}
\nabla_i (B_j \pi^{ij}) & = & 0 \, .
\e
Form these equations we see that a four-dimensional Ricci-flat slice (with $r_{ij} = 0$) can accommodate electrodynamics.
The allowance for this comes from the ghost energy-momentum tensor $\, T^{\mbox{\tiny G}}_{ij} \,$ which fully compensates
the Maxwell's energy-momentum tensor. \\
Consider now the Reissner--Nordstr\o m geometry of a charged particle (see for example, \cite{mtw}):
\b
\label{they}
ds_{(4)}^2  & \!\!\! = \!\!\! &  - (1 - \frac{2 \mu}{r} + \frac{q^2}{r^2}) dt^2 +
(1 - \frac{2 \mu }{r} + \frac{q^2}{r^2})^{-1} dr^2 + r^2 d \Omega^2 \, .
\e
Here $\mu$ is an integration constant called Reissner--Nordstr\o m geometrical mass --- see \cite{som}. It can be
related to the physical mass $m$ and the charge $q$ of the particle via: $\mu = (m^2 + q^2)^{1/2}$. \\
To assure that the four-dimensional part of the solution (\ref{we}) is compatible to the Reissner--Nordst\o m solution
up to and including terms with  $1/r$, we must identify $\alpha$ with $2 \mu$. Hence:
\b
\alpha = 2 \sqrt{m^2 + q^2} \, .
\e
In equation (\ref{we}), $\alpha$ is not the usual Schwarzschild mass but is, instead, the geometrical mass. Then the
integration constant $\beta$ is given by:
\b
\beta = \frac{q}{2 \sqrt{m^2 + q^2}} \, .
\e
The presented example deviates, at fixed $x$, from the Reissner--Nordstr\o m geometry of a charged particle: the
term $q^2/r^2$ has dropped out due to the ghost energy-momentum tensor and this results in to Schwarzschild geometry.
One can therefore interpret the Schwarzschild geometry as arising not from mass only, but from the combined effects of
mass and electric charge; even from electric charge only (for a massless charged centre, $\alpha = 2 \vert q \vert$ and
$\beta = (1/2) \mbox{ sign}(q)$). 

\section{Conclusions}

In conclusion, we would like to point out that the condition $N = \mbox{const}$ is the only ``manually'' imposed condition
in Kaluza's framework. The motivation for this condition is purely physical --- one would expect that Newton's constant
(which is modelled by the dilaton $N$) is indeed a constant. The choice  $N = \mbox{const}$ immediately turns 
equation (\ref{div}) into a gauge-fixing condition for the ``electromagnetic'' potentials $B_i$. \\
Within Kaluza's framework, Einstein's equations are modified by the appearance of extra terms. These terms are called 
{\it ghost} since they are related, as we showed, to the gauge-fixing condition (\ref{e3}) and not to Maxwell's 
equation (\ref{e2}). The modification of Einstein's equation (\ref{e1}) is by no means arbitrary --- it is dictated 
by Kaluza's theory --- and substantially changes the physical picture --- as we showed, a four-dimensional Ricci-flat 
Kaluza Universe can accommodate electrostatics. It is remarkable that the ghost terms are adynamical: the equations of motion for 
$T^{\mbox{\tiny G}}_{ij} =  \nabla^k (\Theta_{ijk} + \Psi_{ijk})$ are trivial in view of (\ref{eqm2}) and (\ref{eqm3}).
Such modification of Einstein's equations brings up the possibility for a cosmological gauge transformation 
of Einstein's tensor $E_{ij}$ in standard general relativity in principle:
\b
\label{cosmol}
E_{ij} \to E_{ij} + \nabla^k \Xi_{ijk} \, ,
\e 
where the extra term $\Xi_{ijk}$, like in Kaluza's theory, is adynamical in leading order of $\kappa$, but not necessarily 
a ghost term (as it is in Kaluza's context). This is also a reminiscence of the fact that the matter energy--momentum tensor 
in flat space is defined modulo an additive divergence of a tensor field \cite{ll}: 
$T^{ij} \to T^{ij} + \partial_k \eta^{ijk} \, ,$ where $\eta^{ijk} = - \eta^{ikj}$ (we have transferred this extra term on 
the ``cosmological side'' of Einstein's equations). In result, Friedmann's equations will be altered (as it happens by the 
introduction of the cosmological constant or in the steady-state cosmology) and issues like dark matter and dark energy 
will fall into new light --- without the restriction $\Omega = 1$. Unlike Kaluza's theory, where the extra terms appear 
naturally, in the general case there would be a freedom to model the extra term $\Xi_{ijk}$ as dictated by experiments --- a 
suitable choice of $\Xi_{ijk}$ could even render general relativity free of dark matter and dark energy. This, however, is 
beyond the scope of the present paper.

\section*{Acknowledgements}

We would like to thank Brian Dolan and Vesselin Gueorguiev for very helpful discussions.

\end{document}